\documentclass[runningheads]{cl2emult}

\usepackage{makeidx}  
\usepackage{graphicx} 
\usepackage{subeqnar} 
\usepackage{multicol} 
\usepackage{cropmark} 
\usepackage{eso}      
\makeindex            

\newcommand{\lesssim}{\mathrel{\raise.3ex\hbox{$<$}\mkern-14mu
             \lower0.6ex\hbox{$\sim$}}}
\newcommand{\gtrsim}{\mathrel{\raise.3ex\hbox{$>$}\mkern-14mu
             \lower0.6ex\hbox{$\sim$}}}
\begin{document}
\title*{The First Stars: \protect\newline
Where did they form? }
\toctitle{The First Stars:
\protect\newline Where did they form? }
%
%
\titlerunning{The First Stars}
%
\author{Jordi Miralda-Escud\'e \inst{1,2}
}
\authorrunning{Jordi Miralda-Escud\'e}
%
%
\institute{University of Pennsylvania,
           David Rittenhouse Laboratory,
           209 St. 33rd St.,
           Philadelphia PA 19104-6396
           jordi@llull.physics.upenn.edu
\and       Alfred P. Sloan Fellow }

\maketitle              

\begin{abstract}
  Several emerging links between high-redshift observational cosmology,
and the Galactic fossil evidence found in the kinematics, metallicities
and ages of Milky Way stars are discussed. In a flat Cold Dark Matter
model with $\Omega\simeq 0.3$ that agrees with present large-scale
structure observations, the oldest stars in the Milky Way should have
formed in the first halos where gas was able to cool, at $z\simeq 20$.
These earliest, weakly bound dwarf galaxies probably turned only a small
fraction of their gas to stars, which should be metal-poor. However, the
merging rate in the early universe was much faster than the present one,
so massive halos with more efficient star formation and metallicities up
to the highest values present today in the bulge could have formed less
than $10^9$ years after the oldest stars. The mean metallicity produced
in the universe by a given redshift is related to the mean surface
brightness of star-forming galaxies above this redshift, and also to the
reionization epoch if galaxies were the dominant sources of ionizing
radiation. The biased distribution of the early dwarf galaxies where the
first stars formed should result in an age gradient with radius of the
low-metallicity stars in the Milky Way, with the oldest ones
concentrated in the bulge and the youngest in the outer halo.
\end{abstract}

\section{Introduction}
  Interest in observations of the Galactic fossil evidence contained in
the spatial distribution and metallicities of stars was for a long time
driven by phenomenological models of the formation of the
Galaxy. The first model \cite{ELS62} proposed a monolithic and rapid
collapse forming the halo first and then the disk. A second one
\cite{SZ78} suggested instead a more gradual collapse of gas clumps to
form the halo. In modern cosmology, the formation of the Galaxy must be
viewed in its cosmological context, where the large-scale structure
theories for the initial fluctuations that collapsed to galaxies are
tested with observations of the Cosmic Microwave Background, galaxy
clustering and evolution, gravitational lensing, Ly$\alpha$ forest
absorption, etc. These ab initio theories should also predict in
principle the statistical characteristics of the formation of a typical
galaxy like the Milky Way (such as the number of clumps that merged to
form our Galaxy at different times and their contents in gas and stars).
In practice, many of these predictions are uncertain due to the large
dynamic range involved, and the complexity of physical processes like
radiative cooling, thermal instability and star formation. We should
nevertheless attempt to identify predictions that circumvent these
uncertainties.

  I review the ideas for galaxy formation in CDM models in \S 2,
discussing the formation of the first stars in \S 3, and where they are
now in \S 4.

\section{Galaxies at $z> 5$}

  I will use here the Cold Dark Matter model with $\Omega=0.3$,
$\Lambda=0.7$, $H_0=65 {\rm km}\, {\rm s}^{-1}\, {\rm Mpc}^{-1}$,
$\Omega_b h^2=0.019$, and normalization $\sigma_8=0.9$. This model seems
to fit basically all available observations of large-scale structure,
\cite{WCOS}, and of Type Ia supernovae as distance indicators
\cite{Pa98,Ra98}.

  In any large-scale structure models hypothesizing the presence
of dark matter that can cluster on small scales, gravitational collapse
starts with low-mass halos, which then grow in mass by merging with each
other and accreting the remaining diffuse matter. The abundances of
halos as a function of redshift is easily visualized in Figure 1, where
the three thick solid lines give the
velocity dispersion (right vertical axis) or virial temperature of the
gas (left vertical axis) of a halo collapsing at redshift $z$ from a
$(1,2,3)-\sigma$ fluctuation, in the model adopted here, calculated
in the Press-Schechter model \cite{PS74,BKCE91}. The lower line
represents the typical object present at a given redshift, while the
upper line gives the most massive objects, containing only about
$0.3$ \% of the mass of the universe if the fluctuations are Gaussian.
This upper line indicates, for example, that at $z=3$ the most
massive clusters should not have velocity dispersions in excess of
$400 \, {\rm km}\, {\rm s}^{-1}$.
The thin dotted lines are of constant halo mass, increasing
in steps of factors of 10.

\begin{figure}
\includegraphics[width=.9\textwidth]{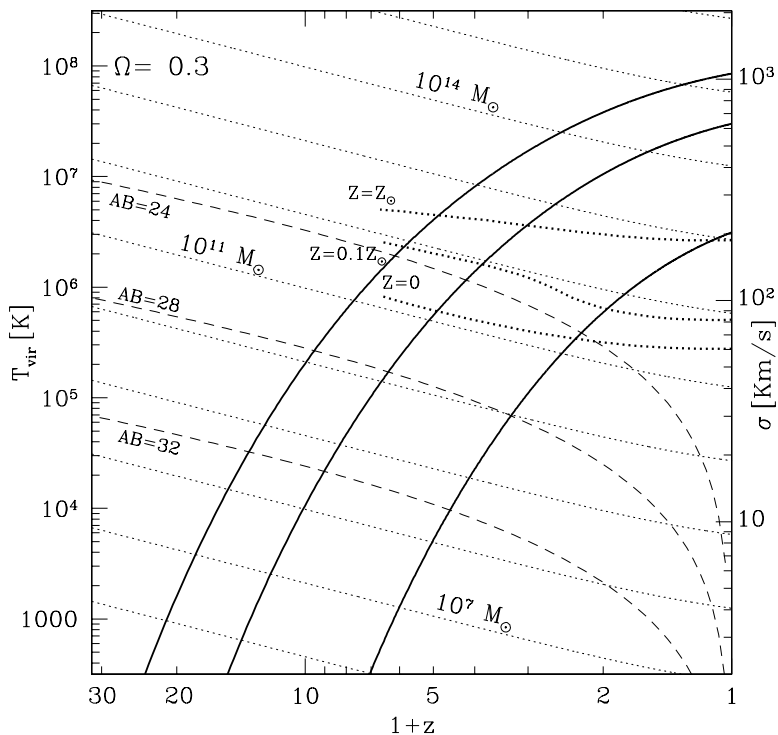}
\caption[]{Velocity dispersion of halos collapsed from $(1,2,3)-sigma$
fluctuations at each redshift ({\it solid thick lines}. See the text for
description of other lines.}
\label{eps1}
\end{figure}

  Also shown as thick dotted lines are the halos where the radiative
cooling time of the gas equals the Hubble time, for three values of the
gas metallicity. Above these lines, the halo gas does not have enough
time to cool, and therefore cannot form galaxies (e.g., \cite{WR78}).
Notice that the gas could still cool in the central parts of these
halos, corresponding to massive clusters, owing to the higher density
near the center; but we know observationally that star formation does
not occur in large amounts in these cooling regions around central
cluster galaxies, for reasons that are not understood \cite{FNC91}.

  After the intergalactic medium was reionized (at $z>5$), galaxies
could only form in halos with $\sigma \gtrsim 30\, {\rm km} \,
{\rm s}^{-1}$, because photoionization heats the gas, raises the Jeans
scale, and makes line-cooling inefficient in objects that collapse only
marginally above the Jeans scale (e.g., \cite{TW96}). Thus, the lower
thick solid line in Figure 1 shows that most of the gas in the universe
had to stay in diffuse form until $z< 2$, which is confirmed in
numerical simulations of the Ly$\alpha$ forest
\cite{MCOR96,ZMAN98}.

  Up to redshifts of $\sim 4$, it was possible to form galaxies with a
total mass similar to a typical $L_{*}$ galaxy today. At that time,
galaxies of this size should have formed by the cooling of all the
gas in some of the most massive halos that existed, and can be
identified with the luminous Lyman break galaxies (e.g., \cite{Sa99}).
But at higher redshifts, the maximum mass of galaxies must rapidly
decline. For example, at $z=8$, galaxies formed from $3-\sigma$
fluctuations should be 100 times less massive than at $z=4$ in Figure 1.
Even if starbursts at high redshift occur on shorter
timescales or are more efficient than at low redshift, the rapid
decline in galaxy mass should most likely cause a similar decline in
galaxy luminosities at $z>5$.

  The mean UV surface brightness of high-redshift galaxies can be
related to the mean metallicity that their stars produced and to the
number of ionizing photons that were emitted \cite{SCL90,MS96,MR98}. We
know that the universe had to be reionized before $z=5$. If star-forming
galaxies were the main sources responsible for reionization, then at
least one ionizing photon per baryon had to be emitted. More likely,
several ionizing photons were needed since most photons could have been
absorbed locally by the interstellar gas in the emitting galaxies, and
the intergalactic gas can also recombine several times before $z=5$. For
every ten ionizing photons emitted per baryon in the universe by stars,
the mean metallicity rises by $10^{-2} Z_{\odot}$, and the galaxies
containing the stars contribute $32$ AB magnitudes per square arc second
to the sky surface brightness in a redshifted UV band to the red of
the Gunn-Peterson trough \cite{MR98}, where galaxies can most easily
be observed at high redshifts. The heavy elements produced by this
first generation of galaxies could be the ones that are observed at
$z\sim 3$ in the diffuse Ly$\alpha$ forest gas \cite{SC96}, and in some
of the lowest metallicity stars we observe at present.

  The dashed lines in Figure 1 indicate the AB magnitude of a starburst
galaxy that forms in a halo of velocity dispersion $\sigma$ at redshift
$z$ (in a rest-frame UV band to the red of the Gunn-Peterson trough),
assuming the optimistic case where all the halo gas turns to stars on
a timescale of the order of the age of the universe (this simple model
is described in more detail in \cite{MR98}). This shows that typical
galaxies at redshift $z=10$ are not likely to be brighter than $AB=30$,
and are probably even fainter if star formation is less efficient
(converting only a small fraction of the halo gas into stars).

\section{The very first stars}

  The first objects where gas is able to cool in the universe correspond
to the high peaks of the density field collapsing on scales of $\sim
10^7 M_{\odot}$, at $z\sim 20$ in the model of Fig. 1 (see \cite{Ba84}).
In halos with temperatures $T\gtrsim 1000$ K the gas cools effectively
through the rovibrational levels of a small fraction of molecular
hydrogen, formed from the primordial ionization fraction via $H^{-}$
\cite{SZ67,PD68,Ta97}. We also see in this Figure that these early halos
merge very rapidly, with the virial temperatures of $3-\sigma$ peaks
increasing from $2000$ K (where molecular cooling is first rapid enough
to lead to the first stars) to 8000 K (where atomic line cooling is already
effective) over the short interval from $z=19$ to $z=16$.

  Recently the first simulations of the formation of the first objects
which compute the formation of the molecular hydrogen, with high enough
resolution to identify the site where the cooled gas concentrates to
form stars, have been performed (\cite{AANZ98,ABN99,N99}. These simulations
show that once the gas cools fast enough, it becomes self-gravitating
and forms a cooling flow in the halo center (notice that because of the
rapid merging rate there should often be several ``halo centers'' or
density peaks in a given collapsed object). Obviously the gas at
the center cools fastest and will collapse first. At the center of the
cooling flow, the cloud becomes optically thick and fully molecular by
three-body reactions, on a time short compared to the evolutionary
timescale of the halo, and this should cause the rapid cooling and
collapse of a central core with a mass of $100$ to $1000 M_{\odot}$
\cite{ABN99,N99}.

  There is much uncertainty about the type of stars that will form
first in the center of this cooling flow \cite{Ta69,KR83,Ca84,S83}:
Will the first stars be massive?
Will many stars be formed by rapid fragmentation?
What differences in the star formation process will be caused by the
absence of heavy elements and magnetic fields? If low-mass stars formed
at the same time as the first massive stars in these low-mass halos,
then there should be some stars with zero metallicity at present;
otherwise, all low-mass stars will have a minimum metallicity generated
by the first supernovae.

  Even if rapid fragmentation led to the formation of a large number of
low-mass stars first, it is clear that massive stars should eventually
form in the first low-mass halos. As long as the formation of stars does
not release much energy, gas will continue to cool and accumulate in the
center, and continued accretion of more gas at the bottom of the
potential cannot be stopped until a sufficiently massive star is formed
to heat and expel the gas around it. A single massive star will produce
about $10^{63}$ ionizing photons, about the same as the number of atoms
in a halo containing $10^6 \, M_{\odot}$ of baryons. Several massive
stars need to form to maintain the entire halo ionized, since the
recombination time at the virial radius of $\sim 100$ pc is typically
$10^5$ years. The evolution after the formation of massive stars is
likely to be very similar to that of molecular clouds in our Galaxy:
their lifetimes are not much longer than $10^7$ years, as they are
destroyed after the formation of a few O stars, turning only 1\% to
10\% of their mass into stars \cite{WM97,WBM99,HT98,CSS95}.
Molecular clouds and the first halos to form stars have similar
baryonic masses, and if anything the first halos are less dense
and their low metallicity should reduce cooling, making them easier to
be heated and dispersed by ionization and supernova explosions compared
to Galactic molecular clouds. Every supernova explosion produces a few
solar masses of metals which, if spread over $10^6 M_{\odot}$ of gas
in a halo, will raise the metallicity by $\sim 10^{-4} Z_{\odot}$.

  Whereas in the Milky Way a destroyed molecular cloud disperses in
the atomic medium, and the gas probably forms molecular clouds again
when it crosses another spiral arm shock \cite{HT98}, in the first
collapsed halos the gas can be expelled in a wind, and can later
collapse again into a more massive object as the sequence of mergers
continues. A self-regulated star formation process can then start,
where new halo mergers produce new starbursts which are stopped when
enough energy is produced by ionizing radiation and supernovae to
disperse the gas again.

\section{Where are the first stars?}

  The self-regulation of star formation in galaxies naturally leads to
the idea that the gas metallicity in a halo will depend mostly on the
velocity dispersion. Gas in weakly bound halos will be heated and
dispersed easily after the formation of a very small number of massive
stars, but in more strongly bound halos star formation can proceed
faster, since the energy output of stars is insufficient to drive a
strong wind, and heated gas is quickly recycled into new molecular
clouds. Since the metallicity depends on the overall fraction of the
gas that has been turned to stars, it should reflect the efficiency of
star formation in each halo.

  The earliest stars formed at $z\simeq 20$ should clearly be
metal-poor. However, in any models where the spectrum of the rms density
fluctuation flattens at small scales (as in CDM), the velocity
dispersions increase very fast at early epochs. Thus, by $z=9$,
$3-\sigma$ peaks (which contain 0.3\% of the mass) are already halos with
$\sigma=50\, {\rm km}\, {\rm s}^{-1}$,
similar to the LMC, which could therefore have
characteristic metallicities $\sim 0.1 Z_{\odot}$, and solar
metallicities can generally be produced as soon as sufficiently massive
objects have collapsed in any model \cite{GO97,CO99}. At the same time, stars
with the lowest metallicities will not all form at the highest redshifts:
most of them will form when the low velocity dispersion halos collapse
from $1-\sigma$ peaks, at $z < 9$. Reionization can in fact stop much of
the gas in the universe from collapsing and forming stars until it
becomes part of a system with $\sigma \gtrsim 30 {\rm km}\, {\rm s}^{-1}$,
implying that most stars at the lowest metallicities would form at $z<2$
for the model of Figure 1.

  Because the age difference between $z=20$ and $z=9$ is only about
400 million years, the prediction that the {\it very oldest} stars
should all be metal-poor will be difficult to test. Until ages for the
oldest stars are not measured with accuracies greater than $\sim 10^9$
years, we should expect essentially no correlation between age and
metallicity \cite{W99}.

\begin{figure}
\includegraphics[width=.8\textwidth]{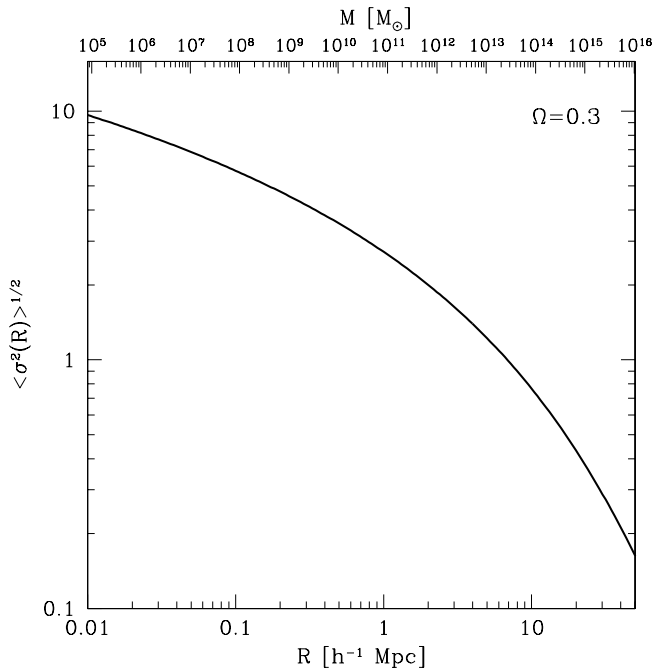}
\caption[]{Rms density fluctuation linearly extrapolated to the present
time on spheres of radius $R$ and mass $M$. }
\label{eps2}
\end{figure}

  Another interesting feature of CDM-like theories is the strong bias
of collapsed objects on small scales, which should cause the oldest
low-metallicity stars to be more abundant in the central parts of
galaxies and clusters.
Figure 2, showing the rms density fluctuation (linearly extrapolated to
the present) as a function of the smoothing scale for the same
CDM$\Lambda$ model used in Fig. 1, serves to illustrate the idea. We
consider a $10^7 M_{\odot}$ halo collapsing from a $3-\sigma$ peak at
$z=18$ as the typical site where the oldest stars were formed. On
average, $3-\sigma$ peaks contain 0.3\% of the mass; however, this
fraction should change in different environments, being high within a
massive galaxy or cluster that collapsed early, and low in a low-mass
galaxy that collapsed late, owing to the bias effect \cite{W99}. As an
example, consider the condition that the halo of the Milky Way has
recently reached a mass of $10^{12} M_{\odot}$, collapsing from a
$1-\sigma$ fluctuation on this scale. Now, we find from Figure 2 that
the ratio of the rms fluctuation on the scales of $10^{12} M_{\odot}$
and $10^7 M_{\odot}$ is $0.3$, so the overdensity due to a $1-\sigma$
fluctuation on $10^{12} M_{\odot}$ already provides 10\% of the
total overdensity needed for a $3-\sigma$ peak at the scale of
$10^7 M_{\odot}$. At the same time, the rms fluctuations on scales
$10^7 M_{\odot}$ that are independent of the smoothed fluctuations on
$10^{12} M_{\odot}$ are reduced by a factor $1-(1/3)^2 = 8/9$. Thus,
given the condition of being part of the Milky Way halo at present,
the abundance of objects of $10^7 M_{\odot}$ formed at $z=18$ should
actually correspond to a density peak $(3-0.3)\cdot (9/8)^{1/2} = 2.86$
times the rms, or an abundance about 50\% higher than the average.
On the other hand, in a massive galaxy collapsing on the same scale of
$10^{12} M_{\odot}$ from a $3-\sigma$ peak at $z=5$ (likely to be part
of a cluster today), the abundance of the $10^7 M_{\odot}$ objects
collapsed at $z=18$ should correspond to $(3-0.9)\cdot (9/8)^{1/2} =
2.23-\sigma$ peaks, or an abundance $\sim 10$ times higher than the
average.

  Of all the galaxies that formed within the region that has collapsed
to the Milky Way today, the ones that ended up merging into the bulge
were the ones that formed in a region with a high overdensity on the
scale of the bulge mass (this assumes that the bulge collapsed before
the Galactic disk was formed, which is supported by the existence of
bulge stars with much lower metallicities than the minimum disk
metallicities). Therefore, they should also contain the largest
abundance of the oldest low-metallicity stars, due to the same
bias effect discussed above. On the other hand, the low-metallicity
stars in the Local Group dwarfs, and in the outer parts of the Milky
Way halo (if they resulted from the disruption of satellite dwarfs)
obviously did not merge into a massive galaxy until they fell recently
to the Milky Way halo, so they were formed in underdense regions on the
scale of the bulge mass, and they are likely to have formed from low
peaks at low redshift. They should consequently have low-metallicity
stars that are younger than in the bulge of the Milky Way and M31.

  This prediction is uniquely characteristic of large-scale structure
theories, since in earlier models it was assumed that the
low-metallicity stars formed in the early generations that produced
the heavy elements in the bulge would still be found in the outer halo,
and would therefore be the oldest stars,
with the gas being enriched as it collapsed further toward the center.
These predictions are discussed here at a very qualitative level, but
improved numerical simulations of galaxy formation combined with
chemical-enrichment models are likely to make these more quantitative
in the near future.

\clearpage
\addcontentsline{toc}{section}{Index}
\flushbottom
\printindex

\end{document}